\documentclass[12pt,aps,preprint,prd,nofootinbib,groupedaddress]{revtex4-1}
\usepackage{hyperref}
\usepackage{amssymb}
\usepackage{amsmath}
\usepackage{bm}
\usepackage{graphicx}

\begin{document}
\title{Non-conformal evolution of magnetic fields during reheating}
\author{Esteban Calzetta}
\email[E-mail me at: ]{calzetta@df.uba.ar}
\affiliation{Departamento de F\'\i sica and IFIBA, FCEyN - UBA, 
Ciudad Universitaria, CABA, Argentina.}
\author{Alejandra Kandus}
\email[E-mail me at: ]{kandus@uesc.br}
\affiliation{LATO - DCET - UESC - Rodovia Ilh\'{e}us-Itabuna km 16 s/n,\\
 Ilh\'{e}us - BA, Brazil}

\begin{abstract}

We consider the evolution of electromagnetic fields coupled to conduction currents during the reheating era
after inflation, and prior to the establishing of the proton-electron plasma. We assume that the currents may 
be described by second order causal hydrodynamics. The resulting theory is not conformally invariant. The expansion 
of the Universe produces temperature gradients which couple to the current and generally oppose Ohmic dissipation. Although 
the effect is not strong, it suggests that the unfolding of hydrodynamic instabilities in these models may follow 
a different pattern than in first order theories, and even than in second order theories on non expanding backgrounds.
\end{abstract}

\maketitle

\section{Introduction}\label{intro}
The existence of magnetic fields on galactic and larger scales is one of the main puzzles in present day cosmology \cite{KKT11,DurNer13,KleFle15}. 
Neither of the two major paradigms proposed to attack this question, namely dynamo amplification and primordial origin, seems to be able to provide 
a solution by itself \cite{FuyYok14}. It therefore seems likely that both mechanisms are at work, i.e., a seed field is generated early in 
the cosmic evolution and then subjected to one or several amplification stages \cite{CT-AK}. This calls for a careful analysis of the cosmological 
history of magnetic fields \cite{BanJed04}. 

Lots of efforts have been made to understand the evolution of primordial fields in the proton-electron plasma during the radiation
dominated epoch. Special mention deserves the studies that address turbulent evolution, where fields with non-trivial topology
i.e., with non zero magnetic helicity, would not be washed out by expansion as quickly as those with null magnetic helicity
\cite{BanJed04,AK-EC,Sigl}.

If we accept the existence of Inflation, then there must be a stage between it and the establishing of the proton-electron plasma
where non-equilibrium processes dominated. That epoch is known as `reheating'. Moreover, electroweak (EW) and quantum-chromodynamic 
(QCD) phase transitions could have taken place by the end of it. Little is known of this epoch, besides the fact that all matter is 
created by the oscillatory decay of the inflaton. For example, the typical relaxation times and correlation times of the different 
interactions are not known.

In this paper we shall perform a preliminary (see below) analysis of the evolution of magnetic fields during the reheating era \cite{AHKK14,MRV14}. 
To this end, we shall consider that, on top of the two dominant contributions to the energy density, namely the coherent oscillations of the 
inflaton \cite{Star80} and the incoherent radiation field, there is a charged fluid that may interact non-trivially with the electromagnetic field. 
We do not identify this fluid with the usual proton-electron plasma because we consider the evolution during an epoch well before 
quantum-chromodynamic phase transition. 

Both the coherent electromagnetic fields and the charged fluid could be created as a side effect of reheating by the parametric amplification of 
vacuum fluctuations of a massive scalar field, as it has been discussed elsewhere \cite{AK-EC}. A suitable candidate for the massive field could 
be the lightest supersymmetric partner, the $s$-$\tau$ \cite{wag-14}. We shall assume that this fluid supports both viscous stresses and conduction 
currents, namely, electric currents without mass transport. For simplicity, we shall use Maxwell theory to describe the fields, in spite of the 
fact that the temperatures involved may be above the electroweak transition.

At those early epochs the temperature and curvature of the Universe are very high and consequently a generally relativistic treatment 
is mandatory.
The theory of relativistic real fluids has a long history but only relatively recently it has been put to the test, through its application to 
relativistic heavy ion collisions (RHICs)\cite{RHICs}. Simply put, the straightforward covariant generalization of the Navier-Stokes 
equations leads to the so-called first order theories (FOTs), of which the Eckart \cite{Eck40} and Landau-Lifshitz \cite{LL6} formulations 
are the best known. These theories have severe formal problems \cite{HisLin83}  which may be solved (among several possible strategies 
\cite{sps}) by going over to the so-called second order theories (SOTs)\cite{SOTs}. The performance of SOTs with respect to RHICs is analyzed in 
\cite{SOT-RHICs}.

There is not a single SOT framework as compelling as the Navier-Stokes equations in the non-relativistic regime \cite{SOTs,extended,dtts}. 
However, in the linearized regime they all agree in providing a set of Maxwell - Cattaneo equations \cite{MaxCat} for the viscous stresses and 
conduction currents, while they differ in the way the transport coefficients in these equations are linked to the underlying kinetic theory 
description \cite{Jaiswal,Kunihiro,Denicol,BRSSS,TakInu10,HuaKoi12,Strickland}. For this reason in this paper we shall consider only the linearized 
regime. This is what makes our analysis preliminary, because it is likely that the most important effects of the fluid - field interaction will be 
connected to nonlinear phenomena such as inverse cascades \cite{BKT14,BerLin14}, field - turbulence interactions \cite{Sigl} and hydrodynamic 
instabilities \cite{inst}. However, as we shall see, already in the linearized regime there are significant qualitative differences between SOTs 
and FOTs, and between SOTs on flat and expanding backgrounds.

There is a large literature on cosmological models based on SOTs \cite{BNK79,PJC82,PavZim,Mar95}. This literature focused for the largest part 
on homogeneous models, where the interest was in how viscous effects modified the cosmic expansion and contributed to entropy generation. 
These analysis showed that there are meaningful differences between ideal, first and second order theories even at the largest scales. 
To our knowledge, the application of SOTs to inhomogeneous models is less developed than FOTs \cite{BBMR14,FTW14}. This consideration also 
contributed to make an analysis such as this paper a necessary first step. We note that a family of exact solutions for the Boltzmann equation 
in expanding backgrounds with a well defined hydrodynamic limit is known, which provides a helpful test bench for the theory \cite{exact}.

In summary, we shall adopt the so-called divergence type theory supplemented by the entropy production variational principle (EPVP) as a 
representative SOT, but will regard the transport coefficients as free parameters, rather than attempting to derive them from an underlying 
kinetic description \cite{EC-PR}. For this reason, our analysis is relevant to any SOT model. 

The equations of the model are the conservation laws for energy - momentum and charge, the Maxwell equations, and the Maxwell - Cattaneo equations
providing closure; for a detailed derivation see \cite{EC14}. In the linearized regime, these equations decouple in three sets of modes, 
sound waves, incompressible shear waves, and electromagnetic waves coupled to conduction currents. We shall consider only the latter.

We shall model the Universe during reheating as a spatially flat Friedman - Robertson - Walker (FRW) model , whose metric in conformal time is 
$ds^2 = a^2\left(\eta\right) \left[ -d\eta^2 + dx^2 + dy^2 + dz^2\right]$, $a\left(\eta\right)$ being the conformal factor. We shall assume 
for the fluid an equation of state $p=\left( 1/3\right) \rho$ and vanishing bulk viscosity. Under these prescriptions, FOTs lead to conformally 
invariant equations \cite{Sigl}. Therefore the electromagnetic fields are suppressed by a $a^{-2}$ factor, on top of the hydrodynamic evolution.
We shall consider the evolution of electromagnetic fields in an environment where the temperature is higher than the QCD
phase transition temperature, i.e. a scenario where SOTs seem to correctly describe the state of the matter.

Unlike FOTs, the equations derived from SOTs are not conformally invariant: the expansion of the Universe creates temperature gradients which 
couple to the fluid velocity and conduction currents. This leads to a weaker suppression of the magnetic fields than expected from a FOT 
framework. This is the main conclusion of this paper. The effect is not large, but suggests that these SOTs models may be more sensitive to 
nonlinear effects, such as hydrodynamic instabilities, than FOTs or even SOTs on non-expanding backgrounds. This possibility will be investigated 
elsewhere.

The paper is organized as follows: In Section \ref{FE} we introduce the formalism and the covariant equations of second order 
hydrodynamics. We analyze the conformal invariance of the theory and derive the equations for the fields as well as for 
the viscous stress and conduction current, showing that the latter are explicitly non conformally invariant. In Section \ref{LI} 
we linearize the equations and propose a simple, toy model, to solve them. In Section \ref{k0} we consider the homogeneous case $k=0$, 
that permits to study the electric field and conduction current separately  from the magnetic field. In section \ref{kl} we consider  
super-horizon modes of astrophysical interest, i.e., $k\ll 1$, and find that the magnetic fields evolves in a way clearly different 
than in FOT's models. In Section \ref{Con} we summarize and discuss our results.
We leave for the Appendix \ref{apa} the analysis of sub-horizon modes $k\gg 1$ as they are not as astrophysically 
interesting as super-horizon modes. In Appendices \ref{apb} and \ref{apc} we quote some secondary results and technicalities for the reader
interested in those details. 
We work with signature $(-,+,+,+)$ and natural units $\hbar = c = k_B = 1$, thus time and length have dimensions of
$energy^{-1}$, while wavenumbers, mass and temperature units are those of $energy$.

\section{General Relativistic Fluid Equations}\label{FE}

\subsection{The equations in covariant form and their {\boldmath $3+1$} decomposition}

We consider a system composed by a neutral plasma plus electromagnetic field in a flat FRW universe, whose
metric in conformal time is $ds^2 = a^2\left(\eta\right) \left[ -d\eta^2 + dx^2 + dy^2 + dz^2\right]$, 
$a\left(\eta\right)$ being the conformal factor. This form of the metric is obtained from the one written in
physical time $t$ by defining $d\eta = H_0dt/a(t)$, with $H_0$ the Hubble constant during Inflation\footnote{With this
definition, $\eta$ is already dimensionless.}. 
If for Inflation we consider the de Sitter prescription, 
$a_I(t) = \exp\left( H_0 t\right)$,  then $a_I\left(\eta\right) =
1/\left( 1 - \eta\right) $ with $\eta \leq 0$.
If for reheating we accept  that during that period the Universe evolves as if it were 
dominated by matter \cite{Star80}, then $a_R(t) = (1+(3/2)H_0t)^{2/3}$ and consequently $ a_R\left(\eta\right) = \left(1 +\eta/2\right)^2 $.
Observe that we have matched the two expressions at $t=\eta = 0$ such that $a_I(0) = a_R(0) = 1$. As $H_0$ is a 
fixed, characteristic energy scale, we can use it to build non-dimensional quantities, as we did with conformal
time, e. g. we define dimensionless lengths and corresponding wavenumbers as $l = H_0\ell$ and $k = \kappa/H_0 $.
Magnetic and electric field units are $energy^2$ so we write $B = \mathcal{B}/H_0^2$ and $E = \mathcal{E}/H_0^2$.
To complete, we quote the temperature $T = \mathcal{T}/H_0$ and the electric conductivity $\Sigma_c = \sigma_c/H_0$.
We use greek letters to denote space-time indices, and latin letters when we deal with spatial-only components.
Besides, we use semicolons to express covariant derivatives and commas to denote partial derivatives; in
particular a 'prime' will denote partial derivative with respect to conformal time, i.e., $A^{\prime}= \partial A/\partial \eta $
To evaluate the different covariant derivatives we need the 
Christoffel symbols, $\Gamma_{\mu\nu}^{\alpha}$ whose only non-null components are $\Gamma_{00}^{0} = a^{\prime}/a$,
$\Gamma_{ij}^{0} = a^{\prime}/a\delta_{ij}$ and $\Gamma_{0j}^{i} = a^{\prime}/a\delta^{i}_{j}$. 

Let $u^{\mu}$ be the fluid four-velocity.  We decompose it as
\begin{equation}
u^{\mu} = \gamma\left(U^{\mu} + v^{\mu}\right)\label{a1}
\end{equation}
with $\gamma = \sqrt{1 - v^2}$. It is satisfied that $u^{\mu}u_{\mu} = -1$ and $U^{\mu}v_{\mu} = 0$. 
$U^{\mu}$ is the velocity of fiducial observers and $v^{\mu}$ represents deviations from Hubble flow, i.e.
peculiar velocities. Each of these velocities defines a congruence of time-lines, for which there is an
orthonormal space-like surface defined through the projectors
\begin{equation}
h^{\mu\nu} = g^{\mu\nu} + u^{\mu}u^{\nu} , ~~~ \Delta^{\mu\nu}= g^{\mu\nu} + U^{\mu}U^{\nu} \label{a1-b}
\end{equation}
The matter is described by the energy momentum tensor, $T^{\mu\nu}$, which we decompose as
\begin{equation}
T^{\mu\nu} = T^{\mu\nu}_0 + \tau^{\mu\nu}\label{a2-a}
\end{equation}
with
\begin{equation}
T^{\mu\nu}_0 = \left(\rho + p\right) u^{\mu}u^{\nu} + pg^{\mu\nu} \label{a2-b}
\end{equation}
and
\begin{equation}
\tau^{\mu\nu} = \frac{2}{15}\tau\mathcal{F}_4\zeta^{\mu\nu}\label{a2-c}
\end{equation}
the viscous stress tensor. In eq. (\ref{a2-c}), $\tau$ is a characteristic relaxation time and $\zeta^{\mu\nu}$ is a Lagrange
multiplier whose evolution equation will be given below; for $\tau \rightarrow 0$ it reduces to the FOT dissipative shear
viscous tensor.
We write the electromagnetic field tensor $F^{\mu\nu}$ in $3+1$ form relative to the fiducial observers as
\begin{equation} 
F_{\mu\nu}= A_{\mu , \nu} - A_{\nu , \mu} = U_{\mu}E_{\nu} - E_{\mu}U_{\nu} + \eta_{\mu\nu\alpha\beta}U^{\beta}B^{\alpha}
\label{a3}
\end{equation}
with $\eta^{0123}= \left[det\left(-g_{\mu\nu} \right)\right]^{-1/2}$. For future use, we define
$\varepsilon^{\mu\nu\alpha} = \eta^{\mu\nu\alpha\beta}U_{\beta}$.
Observe that the electric and magnetic fields are obtained from (\ref{a3})
as $E^{\mu} = F^{\mu\nu}U_{\nu}$ and $B^{\mu} = (1/2) \eta^{\mu\nu\alpha\beta}U_{\nu}F_{\alpha\beta}$
respectively. The electric current is
\begin{equation}
J^{\mu} = \rho_q u^{\mu} + \Upsilon^{\mu}\label{a4}
\end{equation}
with
\begin{equation}
\Upsilon^{\mu} = \frac{e^2}{3}\tau\mathcal{F}_2\zeta^{\mu}\label{a4-b}
\end{equation}
where $\zeta^{\mu}$ is another Lagrange multiplier whose evolution equation is also given below, and that for $\tau \rightarrow 0$ 
gives the usual Ohm's law.

Although we shall regard  $\mathcal{F}_n$ in eqs. (\ref{a2-c}) and (\ref{a4-b}) as free parameters, we observe that
these equations may be derived from a linearized Boltzmann equation \cite{EC14}, in which case they are seen to be
\begin{equation}
\mathcal{F}_n = \int Dp\frac{f_0}{F}\left\vert -u^{\lambda}p_{\lambda}\right\vert^n \label{a5}
\end{equation}
with $f_0$ the one particle distribution function, $Dp=\left( 2\pi\right)^{-3/2}2d^4p \delta\left(p^{\mu}p_{\mu} - m^2\right)$ 
the integration measure ($m$ is the mass of the plasma particles), and where $F$ is a multiplicative factor in the linearized
collision integral. Common choices for $F$ are Marle's prescription \cite{marle1,marle2}, i.e. $F=const.$, and the Anderson-Witting 
proposal \cite{and-witt1,and-witt2} whereby $F=\left\vert u^{\mu}p_{\mu}\right\vert$. 

Observe that in eq. (\ref{a3}) we defined the electromagnetic field relative to fiducial observers. 
It is also with respect to this velocity that we shall define the `total time derivative' or `dot derivative', namely
$\dot A_{\mu} = A_{\mu ;\nu}U^{\nu}$. The `total spatial derivative' is accordingly defined as $ A_{\mu ;\nu}\Delta^{\nu}_{\alpha} $.

The equations we have to solve are the conservation equations (matter coupled to the electromagnetic field plus charge conservation), 
Maxwell equations and two equations that describe the evolution of the Lagrange multipliers $\zeta^{\mu\nu}$ and $\zeta^{\mu}$.
The conservation laws are
\begin{equation}
T^{\mu\nu}_{~~;\nu} = -J_{\nu}F^{\mu\nu}\label{a6}
\end{equation}
\begin{equation}
J^{\mu}_{;\mu} = 0\label{a7}
\end{equation}
and Maxwell equations in covariant form read
\begin{equation}
F^{\mu\nu}_{~~;\nu}= -J^{\mu}\label{a8}
\end{equation}
\begin{equation}
\eta^{\mu\nu\rho\sigma}F_{\nu\rho;\sigma} = 0 \label{a9}
\end{equation}

To our purposes the best is to rewrite the previous equations in 3+1 form relative to fiducial observers. This 
is achieved by projecting each set along $U^{\mu}$ and onto its orthogonal surface described by $\Delta^{\mu\nu}$. 
The projection along $U^{\mu}$ is defined as \cite{ellis-73} $T^{\mu\nu}_{~;\nu}U_{\mu} =\left(T^{\mu\nu}U_{\mu}\right)_{;\nu} 
- T^{\mu\nu}U_{\mu;\nu}$ and the one onto the orthogonal surface as $T^{\mu\nu}_{~;\nu}\Delta_{\mu}^{\alpha}$.
For the set (\ref{a6}) we first replace expression (\ref{a1}) in eq. (\ref{a2-b}) and
define
\begin{eqnarray}
\check{\rho} &=& \gamma^2 \left(\rho + p\right) - p \label{f7}\\
\check p &=& = p + \frac{1}{3}\left(\gamma^2 -1\right)\left(\rho + p\right) \label{f8}\\
\check q^{\mu} &=& \gamma^2\left(\rho + p\right) v^{\mu}\label{f9}\\
\check{\pi}^{\mu\nu} &=& \gamma^2\left(\rho + p\right) v^{\mu}v^{\nu}
-\frac{1}{3}\left(\gamma^2 -1\right)\left(\rho + p\right)\Delta^{\mu\nu}\label{f10}
\end{eqnarray}
We thus write eq. (\ref{a2-a}) as
\begin{equation}
T^{\mu\nu}=\check{\rho}U^{\mu}U^{\nu} + \check p \Delta^{\mu\nu} + U^{\mu}\check q^{\nu} + U^{\nu}\check q^{\mu} +
\check \pi^{\mu\nu} - \frac{2}{15}\tau T^5\zeta^{\mu\nu} \label{a2-d}
\end{equation}
For eqs. (\ref{a4}) plus (\ref{a4-b}) we directly obtain
\begin{equation}
 J^{\mu} = \rho_q\gamma\left(U^{\mu} + v^{\mu}\right) + \frac{e^2}{3}\tau T^3\zeta^{\mu}\label{a7-b}
\end{equation}
To find the evolution equation for the plasma we assume the equation of state $p = \rho /3$. 
For the projection along $U^{\mu}$ of eq. (\ref{a6}) we have
\begin{eqnarray}
 &&\frac{1}{3}\left[\left( 4\gamma^2 - 1\right)\rho\right]_{,\nu} U^{\nu}  + \frac{4}{3}\left(\gamma^2\rho v^{\nu}\right)_{;\nu} 
 + \frac{4}{3}\frac{a^{\prime}}{a^2}\left[\left( 4\gamma^2 - 1\right) \rho \right]
 +\frac{2}{15}\left(\tau T^5\zeta^{\mu\nu}\right)_{;\nu}U_{\mu} \nonumber\\
 =&& E^{\nu}\left(\gamma \rho_q v_{\nu} 
 + \frac{e^2}{3}\tau T^{3}\zeta_{\nu}\right) \label{a6-b}
\end{eqnarray}
while for the spatial projection we obtain
\begin{eqnarray}
&& p_{,\nu}\Delta^{\mu\nu} + \frac{4}{3}\left[\gamma^2\rho 
v^{\alpha}v^{\nu}\right]_{;\nu}\Delta^{\mu}_{\alpha} 
+ 5\frac{ a^{\prime}}{a^2}\check q^{\alpha}\Delta^{\mu}_{\alpha} 
+ \frac{1}{a}\check q^{\alpha\prime}\Delta^{\mu}_{\alpha} \nonumber \\
-&& \frac{2}{15}\tau^{\prime} T^5\zeta^{\alpha 0}\Delta^{\mu}_{\alpha}
- \frac{2}{3}\tau T^{4} T_{,\nu}\zeta^{\alpha\nu}\Delta^{\mu}_{\alpha}
- \frac{2}{15}\tau T^5 \zeta^{\alpha\nu}_{;\nu}\Delta^{\mu}_{\alpha}\nonumber\\
= && \Delta^{\mu}_{\alpha}\left[\rho_qE^{\alpha} + \frac{1}{a}\tilde\varepsilon^{\alpha\nu}_{~~\rho}B^{\rho}
\left( \rho_q v_{\nu} + \frac{e^2}{3}\tau T\zeta_{\nu}\right)\right] \label{a6-c}
\end{eqnarray}
For eq. (\ref{a7}) using (\ref{a7-b}) we have
\begin{eqnarray}
&&\gamma\rho_{q,\mu}U^{\mu} + \gamma\rho_{q,\mu}v^{\mu}
+ \rho_q \left[ \gamma^2 v^{\alpha}v_{\alpha ;\mu} u^{\mu} + \gamma U^{\mu}_{;\mu} + \gamma v^{\mu}_{;\mu} \right] \nonumber\\
+&&  e^2\tau T^2T_{,\mu}\zeta^{\mu}
+ \frac{e^2}{3}\tau T^3\zeta^{\mu}_{;\mu} = 0\label{a7-c}
\end{eqnarray}
As Maxwell equations are already written in terms of $U^{\mu}$ the projection is straightforward.
For the inhomogeneous Maxwell equations (\ref{a8}) we have
\begin{eqnarray}
E^{\nu}_{;\nu} &=& \rho_q - \frac{e^2}{3}\tau T^3\zeta^{\mu}U_{\mu}\label{a8-b}\\
\Delta^{\mu}_{\alpha}\dot E^{\alpha} &=& -2\frac{a^{\prime}}{a^2}\Delta^{\mu}_{\nu} E^{\nu} 
+ \Delta^{\mu}_{\alpha}\eta^{\alpha\nu\rho\sigma}U_{\sigma}B_{\rho ;\nu} - \Delta^{\mu}_{\alpha} J^{\alpha} \label{a8-c}
\end{eqnarray}
while for the homogeneous ones (\ref{a9}) we obtain
\begin{eqnarray}
B^{\beta}_{;\beta} &=& 0\label{a9-b}\\
\frac{1}{a}\Delta^{\mu}_{\gamma}\varepsilon^{\gamma\beta}_{~~\alpha}E^{\alpha}_{;\beta} 
+ 2\frac{a^{\prime}}{a^2}\Delta^{\mu}_{\gamma} B^{\gamma}
+ \Delta^{\mu}_{\gamma} \dot B^{\gamma} &=& 0\label{a9-c}
\end{eqnarray}

We now discuss the equations for the Lagrange Multipliers $\zeta_{\mu}$ and $\zeta_{\mu\nu}$, see \cite{EC-PR} and \cite{EC14} for details.
$\zeta_{\mu}$ and $\zeta_{\mu\nu}$ are transverse with respect to $u^{\mu}$ and $\zeta_{\mu\nu}$ 
is also traceless, i.e. they satisfy
\begin{equation}
\zeta_{\mu}u^{\mu} = 0 = \zeta_{\mu\nu}u^{\mu}, ~~~\zeta_{\mu}^{\mu} = 0 \label{a10}
\end{equation}
Their evolution equations in covariant form are straightforwardly obtained from the corresponding
Minkowski expressions given in Ref. \cite{EC14}. We obtain:
\begin{equation}
\zeta_{\mu} = 2\frac{\mathcal{A}_2}{\mathcal{A}_3}F_{\mu\nu}u^{\nu} - \tau \frac{\mathcal{F}_4}{\mathcal{A}_3}
h^{\alpha}_{\mu}\zeta_{\alpha ;\beta}u^{\beta} - \frac{1}{e^2\mathcal{A}_1}\left(-J^{\alpha}u_{\alpha}\right)_{;\beta}
h^{\beta}_{\mu}\label{a11}
\end{equation}
\begin{eqnarray}
\mathcal{A}_4\left[\zeta_{\mu\nu} + \tau h^{\alpha}_{\mu}h^{\beta}_{\nu}\zeta_{\alpha\beta ;\gamma}u^{\gamma}\right]
&=& \frac{\mathcal{A}_4}{T}\sigma_{\mu\nu} - \tau \mathcal{A}_5\frac{T_{,\beta}u^{\beta}}{T^2}\zeta_{\mu\nu}\nonumber\\
&-& \frac{\mathcal{A}_5}{7}\frac{\tau}{T}\left[u^{\alpha}_{;\alpha}\zeta_{\mu\nu}+\zeta_{\mu\alpha}\sigma^{\alpha}_{\nu}
+\zeta_{\alpha\nu}\sigma^{\alpha}_{\nu}-\frac{2}{3}h_{\mu\nu}\zeta^{(0)\alpha\beta}\sigma_{\alpha\beta}\right]
\label{a12}
\end{eqnarray}
with $\sigma_{\mu\nu} = (1/2)\left[u_{\mu;\nu} + u_{\nu;\mu}\right] - (1/3) u^{\alpha}_{;\alpha}h_{\mu\nu}$ the 
shear tensor. In the derivation from linearized kinetic theory the functions $\mathcal{A}_n$ are given by \cite{EC14}:
\begin{equation}
\mathcal{A}_n = \int Dp\left\vert -u_{\alpha}p^{\alpha}\right\vert^n f_0 \label{a13}
\end{equation}
We only mention this because it makes it easy to check the dimensions of $\mathcal{F}_n$ and $\mathcal{A}_n$;
otherwise we shall regard them as free parameters. 
The dimensions of the different expressions under the integrals are
$ \left[ f_0\right] = 1$, $ \left[ Dp\right] = E^2$, $ \left[ u_{\alpha}p^{\alpha}\right] = E$,
with $E$ meaning 'energy' and consequently $\left[\mathcal{A}_n\right] = E^{n+2}$ and
$\left[\mathcal{F}_n\right] = E^{n+1}$. As the only energy scale of the plasma is its temperature,
we rewrite eq. (\ref{a2-b}) as
\begin{equation}
\tau^{\mu\nu} = \frac{2}{15}c_1\tau T^5\zeta^{\mu\nu}\label{a2-e}
\end{equation}
and eq. (\ref{a4-b}) as
\begin{equation}
\Upsilon^{\mu} = \frac{e^2}{3}c_2\tau T^3\zeta^{\mu}\label{a4-c}
\end{equation}
with $c_1, ~ c_2$ dimensionless, $\mathcal{O}(1)$ coefficients.

\subsection{Conformal Invariance}\label{CI}
To analyze conformal invariance we begin by rewriting the coefficients in eq. (\ref{a11}) and (\ref{a12}) as
\begin{equation}
 \frac{\mathcal{A}_2}{\mathcal{A}_3} = \frac{b_1}{T}, ~~~ \frac{\mathcal{F}_4}{\mathcal{A}_3} = b_2, ~~~
 \frac{1}{\mathcal{A}_1} = \frac{b_3}{T^3}, ~~~ \frac{\mathcal{A}_5}{\mathcal{A}_4} = d_1T\label{a15}
\end{equation}
where $b_1,~b_2,~ b_3,~ d_1$ are again numerical, $\sim \mathcal{O}(1)$ coefficients.
Therefore the mentioned eqs. read
\begin{equation}
\zeta_{\mu} = 2\frac{b_1}{T}F_{\mu\nu}u^{\nu} - b_2\tau h^{\alpha}_{\mu}\zeta_{\alpha ;\beta}u^{\beta} - 
\frac{b_3}{e^2T^3}\left(-J^{\alpha}u_{\alpha}\right)_{;\beta}
h^{\beta}_{\mu}\label{a14}
\end{equation}
\begin{eqnarray}
\frac{1}{T}\sigma_{\mu\nu} - d_1\tau \frac{T_{,\beta}u^{\beta}}{T}\zeta_{\mu\nu} &=&
\left[\zeta_{\mu\nu} + \tau h^{\alpha}_{\mu}h^{\beta}_{\nu}\zeta_{\alpha\beta ;\gamma}u^{\gamma}\right]\nonumber\\
&+& \frac{d_1}{7}\tau\left[ u^{\alpha}_{;\alpha}\zeta_{\mu\nu} + \zeta_{\mu\alpha}\sigma^{\alpha}_{\nu}
+\zeta_{\alpha\nu}\sigma^{\alpha}_{\mu} - \frac{2}{3}h_{\mu\nu}\zeta^{(0)\alpha\beta}\sigma_{\alpha\beta}\right]
\label{a16}
\end{eqnarray}
We now transform the different quantities in the model according to
\begin{equation}
 u^{\mu} = \frac{\tilde u^{\mu}}{a} \rightarrow u_{\mu} = a\tilde u_{\mu} ~~~ h^{\mu\nu} = \frac{\tilde h^{\mu\nu}}{a^2}
 \rightarrow h^{\mu}_{\nu} = \tilde h^{\mu}_{\nu} \rightarrow h_{\mu\nu} =  a^2\tilde h_{\mu\nu}, ~~~
 \sigma_{\mu\nu} = a\tilde \sigma_{\mu\nu} \label{a17}
\end{equation}
(and similar rules for $U^{\mu}$ and $\Delta^{\mu\nu}$)
\begin{equation}
 \zeta^{\mu}=\frac{\tilde \zeta^{\mu}}{a^2}\rightarrow \zeta_{\mu} = \tilde\zeta_{\mu}; ~~~~~
 \zeta^{\mu\nu}=\frac{\tilde \zeta^{\mu\nu}}{a^2}\rightarrow \zeta_{\mu\nu} = a^2\tilde\zeta_{\mu\nu}\label{a18}
\end{equation}
\begin{equation}
F_{\mu\nu}=\tilde F_{\mu\nu} \rightarrow F^{\mu\nu}=\frac{\tilde F^{\mu\nu}}{a^4}\label{a19}
\end{equation}
and
\begin{equation}
\rho = \frac{\tilde \rho}{a^4}, ~~~p=\frac{\tilde p}{a^4}, ~~~ \rho_q = \frac{\tilde \rho_q}{a^4},
~~~ J^{\mu} = \frac{\tilde J^{\mu}}{a^4}, ~~~ T = \frac{T_0}{a}, 
~~~ \tau = a\tilde \tau \label{a20}
\end{equation}
Replacing these transformations in eqs. (\ref{a2-d}), (\ref{a4}), (\ref{a2-e}) and (\ref{a4-c}) we find
\begin{equation}
T^{\mu\nu}_0 = \frac{\tilde T^{\mu\nu}_0}{a^6}, ~~~ \tau^{\mu\nu} = \frac{\tilde \tau^{\mu\nu}}{a^6},~~~
J^{\mu} = \frac{\tilde J^{\mu}}{a^4}\label{a21}
\end{equation}
and the set of eqs.  (\ref{a6-b})-(\ref{a6-c}) becomes
\begin{eqnarray}
&&\frac{1}{3}\left( 4\gamma^2 - 1\right) \tilde\rho^{\prime} + \frac{4}{3}\gamma^{2\prime} \tilde\rho
+ \frac{4}{3}\tilde\rho\gamma^{2}_{,j}\tilde v^{j} + \frac{4}{3}\gamma^2\tilde\rho_{,j}\tilde v^{j}
+ \frac{4}{3}\gamma^2\tilde\rho\tilde v^{j}_{,j}
+ 5 \tilde\tau \tilde T^4\tilde T_{,j}\tilde\zeta^{0j} + \tilde\tau \tilde T^5\tilde\zeta^{0j}_{,j}\nonumber\\
=&& \tilde E^{j}\left[\gamma\tilde\rho_q\tilde v_{j} + \frac{e^2}{3}\tilde\tau\tilde T^3\tilde\zeta_{j}\right]
\label{a6-d}
\end{eqnarray}
\begin{eqnarray}
&& \frac{1}{3}\tilde \rho_{,i} 
+ \frac{4}{3}\left[\gamma^2\tilde \rho  \tilde v^{i}\tilde v^{j}\right]_{,j}  
+ \frac{4}{3}\left(\gamma\tilde\rho \tilde v^{i}\right)^{\prime} 
- \frac{2}{15} \tilde\tau^{\prime} \tilde T^5 \tilde\zeta^{i 0}
-\frac{2}{15} \frac{a^{\prime}}{a}\tilde\tau \tilde T^5 \tilde\zeta^{i 0}
- \frac{2}{3}\tilde\tau \tilde T^{4} \tilde T_{,j}\tilde\zeta^{ij}
- \frac{2}{15}\tilde\tau \tilde T^5 \tilde\zeta^{ij}_{,j}
\nonumber\\
= && \left[\tilde\rho_q\tilde E^{i} + \tilde\varepsilon^{ij}_{~~k}\tilde B^{k}
\left( \tilde\rho_q \tilde v_{j} + \frac{e^2}{3}\tilde\tau  \tilde T^3 \tilde\zeta_{j}\right)\right] \label{a6-e}
\end{eqnarray}
while for eq. (\ref{a7-c}) we have
\begin{equation}
\gamma\tilde\rho_{q}^{\prime}  
+ \gamma\tilde\rho_{q,\mu}\tilde v^{\mu}
+ \tilde\rho_q \gamma^3\left[  \tilde v_{\alpha}^{\prime} 
+ \tilde v_{\alpha ,\mu} \tilde v^{\mu}\right] \tilde v^{\alpha} 
+ \tilde\rho_q \gamma \tilde v^{\mu}_{,\mu}
+ e^2\tilde\tau\tilde T^2 \tilde T_{,\mu}\zeta^{\mu} 
+ \frac{e^2}{3}\tilde\tau\tilde T^3\tilde\zeta^{\mu}_{,\mu} = 0\label{a7-d}
\end{equation}
It is a well known result that Maxwell equations are conformally invariant. For the homogeneous equations it is a trivial result, 
and for the inhomogeneous equations it is
directly apparent from the transformation law for $F^{\mu\nu}$ and the last of exprs. (\ref{a21}). Therefore
transforming eqs. (\ref{a8-b})-(\ref{a9-c}) we have
\begin{eqnarray}
\tilde E^i_{,i} &=& \tilde \rho_q + \frac{e^2}{3}\tilde\tau\tilde T^3\tilde\zeta^{0}\label{a8-d}\\
\tilde E^{i\prime}  &=& \varepsilon^{ij}_{~~k} \tilde B^{k}_{ ,j} - \tilde\rho_q\tilde v^i -\frac{e^2}{3}
\tilde\tau \tilde T^3\tilde\zeta^{i}\label{a8-e}\\
\tilde B^i_{,i} &=& 0 \label{a9-d}\\
 \tilde B^{i\prime}  &=& -\varepsilon^{ij}_{~~k}\tilde E^{k}_{,j}  \label{a9-e}
\end{eqnarray}

Notwithstanding, when we apply the above conformal transformations to the evolution equations for
$\zeta^{\mu}$ and $\zeta^{\mu\nu}$ conformal invariance is lost. To see this, we replace $u^{\mu}$ and $F^{\mu\nu}$
from eqs. (\ref{a1}) and (\ref{a3}), and use the conformal transformations defined above to obtain
\begin{equation}
\tilde\zeta_{\mu} = 2\frac{b_1}{\tilde T}\gamma\left[\tilde E_{\mu} + \tilde U_{\mu}\tilde E_{\nu}\tilde v^{\nu}
+ \tilde\varepsilon_{\mu\nu\alpha}\tilde B^{\alpha}\tilde v^{\nu}\right] - \tilde\tau b_2\gamma\tilde h^{\alpha}_{\mu}
\left[\tilde \zeta_{\alpha , 0} - \frac{a^{\prime}}{a}\tilde\zeta_{\alpha}+\tilde v^{\beta}\tilde\zeta_{\alpha ,\beta}
\right] \label{a14-b} 
\end{equation}
and
\begin{eqnarray}
&&\left[1 + \tilde\tau d_1\gamma \left(\frac{ \tilde T^{\prime}}{\tilde T} 
-\frac{a^{\prime}}{a}\right)\right]\tilde \zeta_{\mu\nu} =
\frac{1}{\tilde T}\tilde \sigma_{\mu\nu} 
- \tilde \tau h^{\alpha}_{\mu}h^{\beta}_{\nu} \gamma
\left[\tilde \zeta_{\alpha\beta}^{\prime} + \tilde\zeta_{\alpha\beta , j}\tilde v^j \right] 
 \label{a15-b}\\
-&& \tilde\tau\frac{ d_1}{7}\left[ \gamma^{\prime} 
+ \gamma\left( 3\frac{a^{\prime}}{a} + \tilde v^{\alpha}_{,\alpha}   \right) \right] \tilde\zeta_{\mu\nu}
- \tilde \tau\frac{ d_1}{7}\left[\tilde\zeta_{\mu\alpha}\tilde\sigma^{\alpha}_{\nu}
+\tilde\sigma_{\mu}^{\alpha}\tilde\zeta_{\alpha\nu} 
- \frac{2}{3}\tilde h_{\mu\nu}\tilde\zeta^{(0)\alpha\beta}\tilde\sigma_{\alpha\beta}\right]\nonumber
\end{eqnarray}
In both equations, the terms proportional to $a^{\prime}/a$ do not cancel out and this fact makes the two
equations non conformal invariant. As the fields evolve coupled to this plasma, the conservation of the magnetic 
flux during their early evolution is lost. To have a glimpse of the effect of this coupling on the amplitude
of the magnetic field, we shall solve the equations in the linear regime.

\section{Linear Evolution}\label{LI}
The system of equations that describe the evolution of the plasma is non linear. We shall study the linear regime, 
that is suitable for small amplitudes. We shall also consider that the plasma is neutral, i.e., we assume 
$\tilde\rho_q =\delta\tilde\rho_q = 0$. First order quantities
are $\tilde\zeta^{\mu}, ~\tilde\zeta^{\mu\nu}, ~ \tilde v^{j}, ~ \delta\tilde\rho$ and the electromagnetic field. 
Writing $H\left(\eta\right) = a^{\prime}/a$, the linear equations read
\begin{eqnarray}
\delta\tilde\rho^{\prime} &=&- \frac{4}{3}\tilde\rho_0\tilde v^j_{,j} \label{b1}\\
 \tilde v^{i}_{,i} &=& - \frac{1}{4\tilde\rho_0}\delta\tilde\rho_{,i}
 +\frac{1}{10}\frac{\tilde\tau  T^{5}_0}{\tilde\rho_0}\tilde\zeta^{ij}_{~,j}\label{b2}\\
 \tilde\zeta^{\prime}_{ij} &=& \frac{1}{\tilde\tau T_0}\tilde\sigma_{ij} + 
\left[\frac{4}{7}d_1 H\left(\eta\right) - \frac{1}{\tilde\tau}\right]\tilde\zeta_{ij}\label{b3}\\
\tilde\zeta_i^{\prime} &=& \left[ H\left(\eta\right) - \frac{1}{b_2\tilde\tau}  \right]\tilde\zeta_i 
+ \frac{b_1}{b_2\tilde\tau T_0}\tilde E_i\label{b4}\\
\tilde E^{i\prime} &=& \varepsilon^{ij}_{~~k}\tilde B^k_{,j} - \frac{e^2}{3}\tilde\tau T_0^3\tilde\zeta^i\label{b5}\\
\tilde B^{i\prime} &=& -\varepsilon^{ij}_{~~ k}\tilde E^{k}_{,j}\label{b6}
\end{eqnarray}
where we see that at this level the plasma equations have separated from the electromagnetic equations, so
from now on we concentrate only in the latter as our focus is the electromagnetic field evolution.
Before going on, observe that if we set $\tilde\tau \rightarrow 0$ in eq. (\ref{b4}) we have that
$\tilde\zeta_i = \left(b_1/T_0\right)\tilde E_i$. Replacing this expression into Amp\`ere law, eq. (\ref{b5}), the 
factor that multiplies $\tilde\zeta^i$ in the last term of the r.h.s. becomes $(e^2/3)\tilde\tau T_0^2 b_1 E^i$,
and recalling the constitutive relation between electric field and density current, $\tilde J^i = \tilde\sigma_c \tilde E^i$,
we can read the expression for the (commoving) electric conductivity:
\begin{equation}
\tilde\sigma_c = b_1\frac{e^2}{3}\tilde\tau T_0^2\label{b7}
\end{equation}
Observe also, that due to the conformal scalings (\ref{a20}) the physical and commoving electric conductivities are
related in the usual way, i.e. $\sigma_c = \tilde\sigma_c/a$. We then rewrite eq. (\ref{b5}) as
\begin{equation}
\tilde E^{i\prime} = \varepsilon^{ij}_{~~k}\tilde B^k_{,j} - \frac{T_0\tilde\sigma_c}{b_1}\tilde \zeta^i\label{b5-b} 
\end{equation}

To go on we change the time dependence from $\eta$ to $u = (1 +\eta/2)$, whence $d/d\eta = (1/2)d/du$ and $H = 1/u$.
Assuming incompressible evolution and transforming Fourier we get
\begin{eqnarray}
\frac{d\tilde\zeta^{i}\left(\bar k,u\right)}{du}  &=& - 2\left[ \frac{1}{b_2\tilde\tau} - \frac1u\right]
\tilde \zeta^{i}\left(\bar k,u\right) + \frac{2b_1}{b_2\tilde\tau\tilde T_0}\tilde E^i \left(\bar k,u\right) 
\label{z-1}\\
\frac{d\tilde E^{i\prime}\left( \bar k,u\right)}{du} &=& 2i\varepsilon^{ij}_{~~k} k^j\tilde B^{k}\left( \bar k,u\right)
 -\frac{2T_0\tilde\sigma_c}{b_1}\tilde\zeta^{i}\left( \bar k,u\right) \label{z-2}\\
\frac{d\tilde B^{i} \left( \bar k,u\right)}{du} &=& -2i\varepsilon^{ij}_{~~k}k^j\tilde E^{k} \left( \bar k,u\right)  \label{z-3}
\end{eqnarray}

We shall not attempt to solve system (\ref{z-1})-(\ref{z-2}) numerically, as this would oblige us to stick to a specific range
of parameters. Instead, to have a glimpse of how the system behaves we assume a simple configuration given by
\begin{equation}
\bar k = k\check z,~~~ \tilde B_i =\tilde B_y \check y,~~~ \tilde E_i =\tilde E_x \check x, ~~~\tilde \zeta_i =\tilde \zeta_x \check x
\label{b10}
\end{equation}
Defining the matrices
\begin{equation}
\mathbf{\Lambda} = \left(
\begin{array}{c}
\tilde\zeta_x\\
\tilde E_x\\
\tilde B_y
\end{array}
\right)\label{b11}
\end{equation}
and
\begin{equation}
 \mathbf{\Xi} =
 \left(
\begin{array}{ccc}
\frac{1}{b_2\tau} & -\frac{b_1}{b_2}\frac{1}{\tau T_0} & 0 \\
\frac{\tilde\sigma_c T_0}{b_1} & 0 & ik\\
0 & ik & 0
\end{array}
\right) , ~~~~~
\mathbf{H} = \left(
\begin{array}{ccc}
 \frac{1}{u} & 0 & 0\\
 0 & 0 & 0\\
 0 & 0 & 0
\end{array}
\right)\label{b12}
\end{equation}
the system of equations for the electromagnetic sector can be written in matrix form as
\begin{equation}
\mathbf{\Lambda}^{\prime} + 2\mathbf{\Xi}\mathbf{\Lambda} = 2\mathbf{H}\mathbf{\Lambda}\label{b13}
\end{equation}
where now a 'prime' denotes derivative with respect to $u$, i.e., $\prime = d/du$.
In spite of its simple form, it is rather difficult to solve eq. (\ref{b13}) exactly, except for the homogeneous mode, 
$k=0$. We begin by solving this case and then consider perturbatively the case $k \ll 1$, that 
corresponds to modes well outside the particle horizon as e.g., the galactic scale. The solution for modes $k \gg 1$ is
given in Appendix \ref{apa}.

To appreciate the features of the SOT evolution, it is convenient to keep in mind their behavior in the  $\tilde\tau \rightarrow 0$ limit, 
whereby the model reduces to a FOT. In that case system (\ref{z-1})-(\ref{z-3}) 
plus (\ref{b7}) and model (\ref{b10}) reduces to
\begin{eqnarray}
\frac{\tilde E\left(k,u\right)}{du}  &=& 2ik\tilde B-2\tilde\sigma_c\tilde E\label{v46}\\
\frac{\tilde B^{i} \left(k,u\right)}{du} &=&-2ik\tilde E \label{v47}
\end{eqnarray}
and this (conformally invariant) system can be combined to give a wave equation whose solutions are the exponentials
$e^{-2\gamma_{(\pm )} u}$ with $\gamma_{(\pm)} = \tilde\sigma_c/2 \pm \sqrt{\tilde\sigma^2_c/4 - k^2}$. Observe that
when $k\rightarrow 0$, $\gamma_{(+)}\rightarrow \sigma_c$ and $\gamma_{(-)}\rightarrow 0$. The second solution describes the 
``frozen'' magnetic field, and the first the ``discharge'' of the electric field due to the resistivity of the plasma.
If $k\not=0$ we have the well known pure exponential decay.

\section{Analytic Solution for the Homogeneous Mode { \boldmath $k=0$}}\label{k0}

In the $k=0$ case, eqs. (\ref{b13}) may be solved in closed form. 
We then begin by putting $k=0$ in matrix $\mathbf{\Xi}$ and the r.h.s. of eq. (\ref{b13}) equal to zero. Proposing as solution a time
dependence of the form $\Lambda^i\left(\eta\right) = \Lambda^{(0)i}\exp\left(-2\lambda^{(0)}u\right)$ and imposing that
the determinant of the resulting system be zero we obtain the eigenvalue equation:
\begin{equation}
\lambda^{(0)2}\left(\frac{1}{b_2\tau} -\lambda^{(0)}\right) - \frac{\tilde\sigma_c}{b_2\tau}\lambda^{(0)} = 0\label{c1}
\end{equation}
whose solutions are
\begin{eqnarray}
\lambda^{(0)}_{(0)} &=& 0 \label{z-13-a}\\
\lambda^{(0)}_{(\pm)} &=& \frac{1}{2b_2\tilde\tau}\left( 1 \pm \sqrt{1 - 4b_2\tilde\tau\tilde\sigma_c }\right)\label{c2}
\end{eqnarray}
Observe that there exists a critical relaxation time, $\tilde\tau_c = 1/\left(4b_2\tilde\sigma_c\right)$. Also, and more
importantly, when $\tilde\tau \rightarrow 0$ we have that $\lambda_{(-)} \rightarrow \tilde\sigma_c$ while $\lambda_{(+)}$
blows out. Therefore $\lambda^{(0)}_{(0)}$ and $\lambda^{(0)}_{(-)}$ converge to the roots of the FOT model.
The corresponding eigenvectors are
\begin{equation}
\mathbf{\Lambda}_{(0)}^{(0)} = \left(
\begin{matrix}
0\\
0\\
1
\end{matrix}
\right) , ~~~~ \mathbf{\Lambda}_{(+)}^{(0)} = \left(
\begin{matrix}
 1\\
 \frac{\tilde\sigma_c T_0}{b_1\lambda_{(+)}}\\
 0
\end{matrix}
\right) , ~~~~ \mathbf{\Lambda}_{(-)}^{(0)} = \left(
\begin{matrix}
 1\\
 \frac{\tilde\sigma_c T_0}{b_1\lambda_{(-)}}\\
 0
\end{matrix}
\right) \label{c3}
\end{equation}

To find the solution of the inhomogeneous equation we propose
\begin{equation}
\Lambda^i = a_{(0)}\left(u\right) \Lambda_{(0)}^{(0)} + a_{(+)}\left(u\right) \Lambda_{(+)}^{(0)} 
e^{-2\lambda^{(0)}_{(+)}u}
+ a_{(-)}\left(u\right) \Lambda_{(-)}^{(0)}e^{-2\lambda^{(0)}_{(-)}u} \label{z-17}
\end{equation}
and substitute in eq. (\ref{b13}). For $a_{(0)}\left(\eta\right)$ it is straightforwardly
obtained that $a_{(0)}\left(\eta\right) = const$. Recalling that this coefficient corresponds to $\Lambda^{(0)}$,
and that this eigenvector represents the magnetic field, this means the obvious result that the commoving field
remains constant and consequently the physical magnetic intensity will decay as
$\propto a^{-2}\left(\eta\right)$. The other coefficients satisfy
\begin{eqnarray}
a_{(+)}^{\prime} &=& \frac{\lambda^{(0)}_{(+)}}{\Delta\lambda^{(0)}} \frac{2}{u} 
\left[ a_{(+)} + a_{(-)}e^{2\Delta\lambda^{(0)}u}\right]
\label{z-20}\\
a_{(-)}^{\prime} &=& -\frac{\lambda^{(0)}_{(-)}}{\Delta\lambda^{(0)}} \frac{2}{u} 
\left[ a_{(+)}e^{-2\Delta\lambda^{(0)}u} + a_{(-)}\right]
\label{z-21}
\end{eqnarray}
with $\Delta\lambda^{(0)} = \lambda^{(0)}_{(+)}-\lambda^{(0)}_{(-)}$. System (\ref{z-20})-(\ref{z-21}) can be reduced to 
\begin{equation}
a_{(+)} = -\left[\frac{\Delta\lambda^{(0)}}{\lambda^{(0)}_{(-)}}\frac{u}{2}a_{(-)}^{\prime} 
+ a_{(-)}\right] e^{2\Delta\lambda^{(0)}u}
\label{z-22}
\end{equation}
plus an equation for $a_{(-)}$:
\begin{equation}
a_{(-)}^{\prime\prime} + \left[  2\Delta\lambda^{(0)}-\frac{1}{u} \right]a_{(-)}^{\prime} + 4\frac{\lambda^{(0)}_{(-)}}{u}a_{(-)}=0
\label{z-24}
\end{equation}
which through the change of variable $z=-2\Delta\lambda^{(0)}u$ can be rewritten as a Kummer equation, whose solutions are
the Confluent Hypergeometric functions \cite{ab-steg,olver-10}.
Two linearly independent solutions of this equation are \cite{ab-steg,olver-10}
$a_{(-)}^{(1)}(u) = U\left(2\lambda^{(0)}_{(-)}/\Delta\lambda^{(0)},-1,-2\Delta\lambda^{(0)}u\right)$ 
and  $a_{(-)}^{(2)}(u) = e^{-2\Delta\lambda^{(0)}u}U\left(-1-2\lambda^{(0)}_{(-)}/\Delta\lambda^{(0)},-1,2\Delta\lambda^{(0)}u\right)$. 
For the given parameters, both functions converge for $u\rightarrow 0$ and $u\rightarrow \infty$ \cite{olver-10}.
Therefore we write $a_{(-)}(u) = \alpha a_{(-)}^{(1)}(u) + \beta a_{(-)}^{(2)}(u)$, with $\alpha$ and $\beta$ constants
to be determined by the initial conditions. 

Before analyzing asymptotic behaviors it is important to establish the (conformal) time interval where the evolution takes place.
As said before, we are considering conduction currents, which are likely to be made of the lightest supersymmetry particle $s$-$\tau$.
This means that we are considering times before the establishing of the standard electron-proton plasma, which we can estimate as
being the time of the QCD phase transition. Moreover, we are interested in the final states of the magnetic evolution coupled to
this current, because it would give the initial conditions for the subsequent evolution of the field in the standard proton-electron
plasma.
If we take the standard value of the Hubble constant during Inflation $H=10^{12}$ GeV  and the Planck mass as  $m_{pl} \simeq 10^{19}$ 
GeV, we estimate the temperature at the onset of reheating as $T_{rh}=\sqrt{H m_{pl}} \simeq 10^{15}$ GeV. This plasma cools down 
due to the expansion as $a_R^{-1} = u^{-2}$. The electroweak phase transition took place at a temperature scale of $T_{EW}\sim 10^2$ GeV,
therefore the dimensionless, conformal time elapsed since the onset of reheating can be estimated as $\Delta u_{(R-EW)}\simeq 
\sqrt{T_{RH}/T_{EW}} \simeq 10^6 \gg 1$. Moreover, if we consider the QCD phase transition, for which $T_{QCD}\sim 10^{-1}$ GeV,
then $\Delta u_{(R-QCD)}\simeq \sqrt{T_{RH}/T_{QCD}} \simeq 10^8 \gg 1$. 
Therefore to find the sought initial values for the subsequent evolution in the radiation era, we can safely take the limit 
$u \gg 1 \sim \infty $ throughout.

When $u\rightarrow \infty$ the Confluent Hypergeometric functions can be always approximated as \cite{ab-steg,olver-10} 
$U(a,b,z) \sim z^{-a}$, and as $e^{-2\lambda^{(0)}_{(+)}u} \ll e^{-2\lambda^{(0)}_{(-)}u}$ we get
\begin{equation}
\tilde\zeta\left( u\rightarrow \infty \right) \sim 
 \alpha  \left(-2\Delta\lambda^{(0)}u\right)^{-2\lambda^{(0)}_{(-)}/\Delta\lambda^{(0)}}
e^{-2\lambda^{(0)}_{(-)}u} \label{z23-b}
\end{equation}
and
\begin{equation}
 \tilde E\left( u\rightarrow \infty \right) \sim  \alpha\frac{ T_0}{b_1}\frac{\tilde\sigma_c}{\lambda_{(-)}}
 \left(-2\Delta\lambda^{(0)}u\right)^{-2\lambda^{(0)}_{(-)}/\Delta\lambda^{(0)}}e^{-2\lambda^{(0)}_{(-)}u}
  \label{z24-b}
\end{equation}
We see that $\zeta \propto \left(b_1/T_0\right) E $. Observe that for $\tilde\tau > \tilde\tau_c $ the solution 
becomes oscillatory. This behavior has no analog in FOT's. However, if we take the limit $\tilde\tau\rightarrow 0$ 
in eq. (\ref{z24-b}), where $\lambda^{(0)}_{(-)}/\Delta\lambda^{(0)}\rightarrow 0$ and 
$\lambda^{(0)}_{(-)}\rightarrow \tilde\sigma_c$, we recover the FOT results.

\section{Non-homogeneous Mode {\boldmath $k \ll 1$}}\label{kl}
Although the homogeneous mode has the appeal of affording a full analytical solution, it is clearly not very interesting from the 
cosmological point of view. In this section we shall consider the set of modes which are most relevant to cosmology, namely modes 
which are far beyond the horizon at reheating, $k \ll 1$. Now the magnetic field is no longer decoupled from the electric field, 
and we expect to find some feedback from the latter on the former, eventually reducing the cosmological $a^{-2}$ suppression and 
the exponential decay found in the FOT analysis. We shall not attempt a full solution, but rather analyze the asymptotic 
behavior of the magnetic field.

To solve for $k\not=0$ we begin by solving perturbatively the eigenvalue equation $\det\mathbf{\Xi}=0$.
\begin{equation}
\lambda^{(k)2}\left( \lambda^{(k)} - \frac{1}{b_2\tau}\right) + \frac{\tilde\sigma_c}{b_2\tau}\lambda^{(k)} = 
-\left( \lambda^{(k)} - \frac{1}{b_2\tau}\right)k^2\label{d1}
\end{equation}
by proposing
\begin{equation}
\lambda^{(k)} = \lambda^{(0)} + \lambda^{(2)}k^2 + \cdots \label{d2}
\end{equation}
After replacing (\ref{d2}) into (\ref{d1}) and keeping terms up to order $k^2$ we find
\begin{eqnarray}
\lambda^{(2)}_{(0)} &=& \frac{1}{\tilde\sigma_c}\label{d4}\\
\lambda^{(2)}_{(\pm)} &=& \frac{\lambda^{(0)}_{(\mp)}}{\lambda^{(0)2}_{(\pm)} -\lambda^{(0)}_{(+)}\lambda^{(0)}_{(-)} }\label{d5}
\end{eqnarray}
with $\lambda^{(0)}_{(+)}\lambda^{(0)}_{(-)} = \tilde\sigma_c/(b_2\tilde\tau)$.
To find the eigenvectors we again set to zero the r.h.s. of eq (\ref{b13}), and propose the new eigenvectors as linear
combinations of the $k=0$ ones, i.e.,
\begin{equation}
\mathbf{\Lambda}^{(k)} = a_{(0)}(k)\mathbf{\Lambda}^{(0)}_{(0)} + a_{(+)}(k)\mathbf{\Lambda}^{(0)}_{(+)}
+ a_{(-)}(k)\mathbf{\Lambda}^{(0)}_{(-)} \label{d6}
\end{equation}
The results are shown in Appendix \ref{apb}.
 
To solve the time evolution we rewrite eq. (\ref{b13}) as
\begin{equation}
u\left[\frac{d}{du}{\Lambda}^{i} + 2{\Xi}^i_j{\Lambda}^j\right] = 2\delta^i_1\delta^1_j{\Lambda}^j\label{d10}
\end{equation} 
In the above equation, upper index in $\Xi$ denotes row while lower index denotes column.
Keeping in mind that the physical range of $u$ starts at $u=1$, we Laplace transform ${\Lambda}^{i}$ as
\begin{equation}
F^i(s) = \int_0^{\infty}du ~e^{-su}{\Lambda}^{i}(u)\label{d11}
\end{equation}
and so eq. (\ref{d10}) becomes
\begin{equation}
-\frac{d}{ds}\left[s F^i(s) - {\Lambda}^i(0) + 2{\Xi}^i_jF^j(s)\right] = 2\delta^i_1\delta^1_jF(s)^j \label{d12}
\end{equation}
As the term involving the initial condition vanishes upon deriving we obtain
\begin{equation}
\frac{d}{ds}\left[\Theta (s)^i_jF^j(s)\right] = -2\delta^i_1\delta^1_jF(s)^j \label{d13}
\end{equation}
where we have defined
\begin{equation}
 \Theta(s)^i_j = 2\Xi^i_j + s\delta^i_j \label{d14}
\end{equation}
We now introduce the inverse matrix of $\Theta(s)^i_j$, $M(s)^i_j$, i.e.
\begin{equation}
\Theta(s)^i_jM(s)^j_k = \delta^i_k \label{d15}
\end{equation}
and define a new variable $K(s)^i$ such that
\begin{equation}
 F^i(s) = M(s)^i_j K(s)^j \label{d16}
\end{equation}
Replacing in eq. (\ref{d13}) we obtain the following equation for $K(s)^i$:
\begin{equation}
 \frac{d}{ds}\left[K^i(s)\right] = -2\delta^i_1M(s)^1_k K(s)^k \label{d17}
\end{equation}
We see that for $i = 2,~ 3$ the solutions are constants. For $i=1$ we have
\begin{equation}
 \frac{dK^1(s)}{ds} + 2M^1_1(s) K^1(s) = -2M^1_2(s) K^2 - 2M^1_3(s) K^3 \label{d18}
\end{equation}

Previously, we have solved the eigenvector equation for the homogeneous system, i.e., $\Xi^i_j \Lambda^{(k)j}_{(\alpha)} 
= \lambda^{(k)}_{(\alpha)}\Lambda^{(k)i}_{(\alpha)} $
with $\alpha = 0,~ +,~ -$ (no sum over greek indices). To avoid cumbersome notation from now on
the label $(k)$ is omitted. There exists the inverse matrix to $\Lambda^i_{(\alpha)}$,
$\Pi^{(\alpha)}_i$,  i.e.
\begin{equation}
\Pi^{(\beta)}_j\Lambda^j_{(\alpha)} = \delta^{(\beta)}_{(\alpha)}\label{d20}
\end{equation}
that also satisfies
\begin{equation}
  \sum_{\alpha} \Lambda^i_{(\alpha)}\Pi^{(\alpha)}_j = \delta^i_j \label{d21}
\end{equation}
Using this result we can write $\Xi^i_j$ as
\begin{equation}
\Xi^i_j = \sum_{(\alpha)} \lambda_{(\alpha )} \Lambda^i_{(\alpha)}\Pi^{(\alpha)}_j\label{d22}
\end{equation}
from where we can write the matrix $M^i_j$ as
\begin{equation}
M^i_j(s) = \sum_{\alpha} \Lambda^i_{(\alpha)}\left( s+2\lambda_{(\alpha)}\right)^{-1}\Pi^{(\alpha)}_j\label{d23}
\end{equation}
We now solve eq. (\ref{d18}). The homogeneous solution is straightforwardly obtained and reads
\begin{equation}
K^1_{hom} (s) = \prod_{\alpha}\left(s + 2\lambda_{\alpha}\right)^{-2A_{(\alpha )}}\label{v17}
\end{equation}
with
\begin{equation}
A_{(\alpha )} = \Lambda^1_{(\alpha )}\Pi^{(\alpha )}_1\label{v18}
\end{equation}
(no sum over $\alpha$). Observe that using relation (\ref{d21}), $A_{(\alpha )}$ satisfies 
\begin{equation}
\sum_{(\alpha )} A_{(\alpha )} = 1\label{v20}
\end{equation}

To find the inhomogeneous solution we propose $K^1_I(s) = L(s)K^1_{hom} (s)$ and after substituting in eq.
(\ref{d18}) we find the following evolution equation for $L(s)$:
\begin{equation}
\frac{d}{ds}L(s) = -2\left[ M_2^1(s) K^2 + M_3^1(s) K^3\right]\prod_{\alpha}\left(s + 2\lambda_{\alpha}\right)^{2A_{\alpha}}
\label{v19}
\end{equation}

Up to here, all the developments have been exact. However, to find solutions that represent the resulting field after 
the evolution in the reheating plasma, we must solve (\ref{v19}) in the asymptotic range $s\rightarrow 0$ 
(i.e., $u \rightarrow \infty$).

\subsection{Solutions for {\boldmath $s\rightarrow 0$} ({\boldmath $u\rightarrow \infty$}):}
We now look into the small $s$ limit. We begin by recalling that one of the eigenvalues, $\lambda_{(0)}$ goes to zero as $k\to 0$, 
while the other two $\lambda_{(+)}$ and $\lambda_{(-)}$ remain finite. Therefore for small enough $s$ we can take 
$s\ll\lambda_{(\pm)}$, but we cannot assume $s\le\lambda_{(0)}$. Therefore, retaining this last eigenvalue explicitly, 
we get (up to an unessential constant)
\begin{equation}
K_{hom}^{1}\left(s\right) \approx\left(s+2\lambda_{(0)}\right)^{-2A_{(0)}}\label{v31}
\end{equation}
and 
\begin{equation}
M\left(s\right)^i_k=\left( \left[2\Xi\right]^{-1}\right)^i_k +\Lambda^i_{0}\Pi^{0}_k\left[ \left(s+2\lambda_{(0)}\right)^{-1}
-\left(2\lambda_{(0)}\right)^{-1}\right] 
\label{v32}
\end{equation}
with
\begin{equation}
 \left[2\Xi\right]^{-1} = \left(
 \begin{matrix}
  \frac{b_2\tilde\tau}{2} & 0 & -\frac{ib_1}{2k T_0}\\
  0 & 0 & -\frac{i}{2k}\\
  \frac{ib_2\tilde\tau T_0\tilde\sigma_c}{2b_1 k} & -\frac{i}{2k} & \frac{\tilde\sigma_c}{2k^2}
 \end{matrix}
\right)\label{xi-1}
\end{equation}

The solution of the inhomogeneous equation now reads
\begin{eqnarray}
L\left(s\right)&=&\frac{-2}{2A_0+1}\left\{\left[ \left( \left[2\Xi\right]^{-1}\right)^1_2 -\Lambda^1_{(0)}\Pi^{(0)}_2
\left(2\lambda_{(0)}\right)^{-1}\right] K^{2}\right.\nonumber\\
&+&\left.\left[ \left( \left[2\Xi\right]^{-1}\right)^1_3 -\Lambda^1_{(0)}\Pi^{(0)}_3
\left(2\lambda_{(0)}\right)^{-1}\right] K^{3}\right\}\left(s+2\lambda_{0}\right)^{2A_{0}+1}\nonumber\\
&-&\frac{1}{A_0}\left[\Lambda^1_{0}\Pi^{0}_2K^2+\Lambda^1_{0}\Pi^{0}_3K^3\right]\left(s+2\lambda_{0}\right)^{2A_{0}}+K^{0}\label{v33}
\end{eqnarray}
whereby
\begin{eqnarray}
K^{1}\left(s\right)&=&L\left(s\right)K_{hom}\left(s\right)\nonumber\\
&=&\frac{-2}{2A_0+1}\left\{\left[ \left( \left[2\Xi\right]^{-1}\right)^1_2 -\frac{\Lambda^1_{(0)}\Pi^{(0)}_2}{2\lambda_{(0)}}\right]K^{2}
+\left[\left( \left[2\Xi\right]^{-1}\right)^1_3 -\frac{\Lambda^1_{(0)}\Pi^{(0)}_3}{2\lambda_{(0)}}\right]K^3\right\}
\left(s+2\lambda_{(0)}\right)\nonumber\\
&-&\frac{1}{A_{(0)}}\left[\Lambda^1_{(0)}\Pi^{(0)}_2K^{2}+\Lambda^1_{(0)}\Pi^{(0)}_3K^{3}\right]+K^{0}
\left(s+2\lambda_{0}\right)^{-2A_{(0)}}
\label{v34}
\end{eqnarray}

The different functions then read
\begin{eqnarray}
F^{i} &=& M^i_j(s)K^{j}(s) = \left(\left[2\Xi\right]^{-1}\right)^i_1K^{1}(s) + \left(\left[2\Xi\right]^{-1}\right)^i_2K^{2}(s) +
\left(\left[2\Xi\right]^{-1}\right)^i_3K^{3}(s) \nonumber\\
&+& \left[ \frac{1}{s+2\lambda_{(0)}} - \frac{1}{2\lambda_{(0)}}\right]
\Lambda^i_0\left[\Pi^0_1K^{1} +\Pi^0_2K^{2} + \Pi^0_3K^{3}\right]\label{v35}
\end{eqnarray}
Our main interest is $F^{3}$ as it is directly related to the magnetic field. 
The calculations are long but straightforward and are shown in Appendix \ref{apc}. The result is
\begin{eqnarray}
F^{3} &\simeq & -\frac{\Lambda^3_0\Pi^0_1}{\left(s+2\lambda_0\right)}
\frac{1}{A_{(0)}}\left[\Lambda^1_{(0)}\Pi^{(0)}_2K^{2}+\Lambda^1_{(0)}\Pi^{(0)}_3K^{3}\right]+
\frac{\Lambda^3_0\Pi^0_1}{\left(s+2\lambda_0\right)^{1+2A_{(0)}}}K^{0}
 \nonumber\\
&+&\frac{\Lambda^3_0\Pi^0_3}{\left(s+2\lambda_0\right)} K^3
 + \frac{\Lambda^3_0\Pi^0_2}{\left(s+2\lambda_0\right)} K^2
\label{v433b}
\end{eqnarray}
The calculation of elements $\Pi^i_j$ is also rather long but straightforward, here we quote the one in the term
with $K^{0}$ as this term gives the main contribution. It reads $\Pi^0_1 = ikb_2\tilde\tau T_0/b_1$, and 
we then have
\begin{equation}
F^{3} \simeq  
\frac{b_2\tilde\tau T_0 K^{0}}{b_1\left(s+2\lambda_0\right)^{1+2A_{(0)}}} ik
\label{v44b}
\end{equation}
where $A_{(0)}\simeq b_2\tilde \tau k^2/\tilde\sigma_c > 0$. The corresponding anti-transformed function is
\begin{eqnarray}
B_y^{(k)}(u) &\sim & \left[\frac{b_2\tilde\tau T_0 K^{0}}{b_1} ik u^{2A_{(0)}} + \mathcal{O}\left(\frac{1}{u}\right)\right]
\exp \left[-2\lambda_0 u\right]
\label{v45b}
\end{eqnarray}
Observe that due to the presence of the factor $u^{2A_{(0)}}$ the magnetic field decays slower than the exponential
law of FOTs, even at large times.

\section{Conclusions}\label{Con}
In this paper we have studied the evolution of electromagnetic fields coupled to conduction currents during the reheating era, 
using second order causal hydrodynamics to describe the evolution of the currents. The evolution of the magnetic field
occurs well before the EW phase transition, during an epoch where the standard proton-electron plasma is not established yet; 
the conduction currents we consider are likely to be made of the lightest supersymmetric partner $s$-$\tau$.
The main motivation behind the  choice of SOTs is the well known fact that first order theories 
(as e.g. relativistic Navier Stokes equation) have severe problems of causality and have no stable equilibrium states.
Also, SOTs behave quite well at describing RHICs \cite{RHICs,SOT-RHICs}, where a plasma much like the one in the very early Universe 
is supposed to be created. Thus, although there is not a preferred SOT framework yet, it is important to begin to study different plasma 
effects in the early Universe using those formalisms extended to general relativity. We adopted the so-called divergence type theory plus 
the entropy production variational principle (EPVP) as a representative SOT, but regarded the transport coefficients as free parameters, 
rather than attempting to derive them from an underlying kinetic description \cite{EC-PR}. In this sense our analysis is relevant to any 
SOT model. When extended to General Relativity, we found that the resulting theory is not conformally invariant: the Maxwell-Cattaneo 
equations that describe the viscous stresses and conduction currents lost this symmetry. As these equations are coupled to Maxwell equations, 
the consequence is that the magnetic flux is not suppressed by the expansion as quickly as in the Navier-Stokes theory. This might
provide higher intensities as initial conditions for the subsequent evolution during radiation dominance. The physical explanation is that 
the expansion of the Universe produces temperature gradients which couple to the current and generally oppose dissipation. 

To pursue the analysis we considered only the linear evolution because in this regime all SOTs agree in providing the set of Maxwell - 
Cattaneo equations.  Our goal was to identify the qualitative differences between FOTs and SOTs, this is the reason why we did not attempt 
to give numerical estimates of the resulting amplitudes. We have found that the field decay in the homogeneous mode may be oscillatory. Even 
in the purely decaying regime, for inhomogeneous modes there is a power-like correction to exponential decay, with a positive
exponent. This suggests that the unfolding of hydrodynamic 
instabilities in these models follows a different pattern than in first order theories, and even than in second order theories on non 
expanding backgrounds. The study of the non-linear hydrodynamic instabilities is the next step in the research of primordial magnetic fields 
evolution within SOTs.

\acknowledgments
E.C. acknowledges support from CONICET, UBA and ANPCyT. A. K. thanks the Physics Department of Facultad de Ciencias
Exactas y Naturales - UBA for kind hospitality during the development and completion of this work, 
and also support from UESC, BA-Brasil.

\appendix

\section{Large {\boldmath $k$} modes}\label{apa}

Although small scales are of little astrophysical interest concerning galactic magnetism, for completion we devote this appendix to 
analyze their evolution with the formalism considered in the paper. Moreover, in this case the mathematics is much simpler. We shall see 
that the effect of the relaxation time $\tau$ is to add damping, while the effects of conformal invariance breaking is to add a slight 
amplification of the magnetic field if the temperature (and therefore the conductivity) is low enough. The equations were
\begin{eqnarray}
\frac{d\tilde\zeta}{du}+\left[ \frac 2{b_2\tau}-\frac{2}{u}\right] \tilde\zeta -\frac{2b_1}{b_2\tilde\tau T_0}\tilde E&=&0\label{r1}\\
\frac{d\tilde E}{du}+2ik\tilde B+\frac{2\tilde\sigma_c T_0}{b_1}\tilde\zeta &=&0\label{r2}\\
\frac{d\tilde B}{du}+2ik\tilde E&=&0\label{r3}
\end{eqnarray}
It is convenient to introduce $\mathcal{E}=\tilde E+\tilde B$ and $\mathcal{B}=\tilde B-\tilde E$ to get
\begin{eqnarray}
\frac{d\tilde\zeta}{du}+\left[ \frac 2{b_2\tau}-\frac{2}{u}\right] \tilde\zeta -\frac{2b_1}{2b_2\tilde\tau T_0}
\left( \mathcal{E-B}\right) &=&0\label{r4}\\
\frac{d\mathcal{E}}{du}+2ik\mathcal{E}+\frac{2\tilde\sigma_c T_0}{b_1}\tilde\zeta &=&0\label{r5}\\
\frac{d\mathcal{B}}{du}-2ik\mathcal{B}-\frac{2\tilde\sigma_c T_0}{b_1}\tilde\zeta &=&0\label{r6}
\end{eqnarray}
Now we write 
\begin{eqnarray} 
\mathcal{E}&=&\mathcal{E}_0e^{-i2ku}\label{r7}\\
\mathcal{B}&=&\mathcal{B}_0e^{i2ku}\label{r8}\\
\tilde\zeta&=& \tilde\zeta_{+}e^{i2ku}+\tilde\zeta_{-}e^{-i2ku}\label{r9}
\end{eqnarray}
with the understanding that the pre-exponentials are all slowly varying functions of time. 
Collecting positive and negative frequency oscillations we get 
\begin{eqnarray}
\frac{\tilde\zeta_{+}}{du}+\left[2ik+ \frac 2{b_2\tau}-\frac{2}{u}\right] \tilde\zeta_{+} +\frac{b_1}{b_2\tilde\tau T_0}
\mathcal{B}_0&=&0\label{r10}\\
\frac{d\mathcal{B}_0}{du}-\frac{2\tilde\sigma_c T_0}{b_1}\tilde\zeta_{+} &=&0\label{r11}
\end{eqnarray}
and 
\begin{eqnarray}
\frac{d\tilde\zeta_{-}}{du}+\left[-2ik+ \frac 2{b_2\tau}-\frac{2}{u}\right] \tilde\zeta_{-}
-\frac{b_1}{b_2\tilde\tau T_0}\mathcal{E}_0 &=&0\label{r12}\\
\frac{d\mathcal{E}_0}{du}+\frac{2\tilde\sigma_c T_0}{b_1}\tilde\zeta_{-} &=&0 \label{r13}
\end{eqnarray}
Leading to
\begin{eqnarray}
\frac{d^2\mathcal{B}_0}{du^2}+\left[2ik+ \frac 2{b_2\tilde\tau}\right] \frac{d\mathcal{B}_0}{du} 
+\frac{\tilde\sigma_c}{b_2}B_0&=&\frac{2}{u}\frac{d\mathcal{B}_0}{du}\label{r14}\\
\frac{d^2\mathcal{E}_0}{du^2}+\left[-2ik+ \frac 2{b_2\tilde\tau}\right] \frac{d\mathcal{E}_0}{du} 
+\frac{\tilde\sigma_c}{b_2}\mathcal{E}_0&=&\frac{2}{u} \frac{d\mathcal{E}_0}{du}\label{r15}
\end{eqnarray}
Let us analyze the equation for $\mathcal{B}_0$. Setting the r.h.s. of eq. (\ref{r14}) to zero, 
the solutions are $e^{i\omega u}$ with
\begin{equation}
\omega^2+\left[2k - \frac {2i}{b_2\tau}\right]\omega -\frac{\tilde\sigma_c}{b_2\tilde\tau}=0\label{r16}
\end{equation}
The roots are
\begin{equation}
\omega_{\pm}=\frac12\left[ \pm\sqrt{\left(2k- \frac {2i}{b_2\tau}\right)^2
+\frac{\tilde\sigma_c}{b_2\tilde\tau} }-\left(2k- \frac {2i}{b_2\tau}\right)\right] \label{r17}
\end{equation}
The slowly varying solution being $\omega_+$. Therefore we postulate
\begin{equation} 
\mathcal{B}_0=ae^{i\omega_{+}u}+be^{i\omega_{-}u}\label{r18}
\end{equation}
to get 
\begin{eqnarray} 
\frac{da}{du}e^{i\omega_+u}+\frac{db}{du}e^{i\omega_-u}&=&0\label{r19}\\
\omega_+\frac{da}{du}e^{i\omega_+u}+\omega_-\frac{db}{du}e^{i\omega_-u}&=&\frac{2}{u}
\left(\omega_+ae^{i\omega_+u}+\omega_-be^{i\omega_-u}\right)  
\label{r20}
\end{eqnarray}
which for $b\ll a$ becomes
\begin{eqnarray}
\frac{da}{du}&=&\frac{2\omega_+}{\left(\omega_+-\omega_-\right)u}a\label{r21}\\
\frac{db}{du}&=&\frac{-2\omega_+}{\left(\omega_+-\omega_-\right)u}ae^{i\left(\omega_+-\omega_-\right)u }\label{r22}
\end{eqnarray}
whose solution for $a$ is  
\begin{equation}
a=u^{\alpha} ~~~~\mathrm{with}~~~~\alpha=\frac{2\omega_+}{\omega_+-\omega_-} \label{r23}
\end{equation} 
Therefore we get
\begin{equation}
\mathcal{B}_0\approx\exp\left\lbrace i\omega_+\left[ \frac{-2i}{\omega_+-\omega_-}\ln u+u\right]  \right\rbrace \label{r24}
\end{equation}
When $k$ is very large we have 
\begin{equation}
\omega_+=\frac{i\tilde\sigma_c}{2}\frac{1}{1+ikb_2\tilde\tau}\label{r25}
\end{equation}
so
\begin{equation} 
\mathrm{Re}\;i\omega_+=\frac{-\tilde\sigma_c}{2}\frac{1}{b_2^2\tilde\tau^2k^2+1}\label{r26}
\end{equation}
and
\begin{equation}
\frac{2\omega_+}{\omega_+-\omega_-}=-\frac{b_2\tau\tilde\sigma_c}{\left[ 1+ikb_2\tilde\tau\right] ^2}\label{r27}
\end{equation}
therefore
\begin{equation} 
\mathrm{Re}\;
\frac{2\omega_+}{\omega_+-\omega_-}= b_2{\tilde\tau\tilde\sigma_c}
\frac{b_2^2\tau^2k^2-1}{\left[ {b_2^2\tau^2k^2+1}\right] ^2}\label{r28}
\end{equation} 
We may write
\begin{equation}
\left\vert \mathcal{B}_0\right\vert\approx\exp\left\lbrace \varDelta\left[ u_c\ln u -u\right]  \right\rbrace \label{r29}
\end{equation} 
where
\begin{equation}
u_c=2 b_2\tilde\tau\frac{b_2^2\tilde\tau^2k^2-1}{b_2^2\tilde\tau^2k^2+1}\label{r30}
\end{equation}
\begin{equation} 
\varDelta=\frac{\tilde\sigma_c}{2}\frac{1}{b_2^2\tau^2k^2+1}\label{r31}
\end{equation}
$\mathcal{B}_0$ grows up to $u_c$
with an amplification factor
\begin{equation}
\left\vert\frac{\mathcal{B}_0\left( u_c\right) }{\mathcal{B}_0\left( u_i\right) }\right\vert=
\exp\left\lbrace \varDelta u_c \left[\ln \left(u_c/u_i\right)-1+u_i/u_c \right] \right\rbrace \label{r32}
\end{equation} 
Of course, provided $u_c>u_i$. For all practical purposes, the amplification is
\begin{equation} 
\left\vert\frac{\mathcal{B}_0\left( u_c\right) }{\mathcal{B}_0\left( u_i\right) }\right\vert=\exp\left\lbrace \frac{1 }{b_2\tilde\tau}
\frac{\tilde\sigma_c}{k^2}\right\rbrace \label{r33}
\end{equation}

\section{Eigenvectors for $k\ll 1$}\label{apb}
After long but straightforward calculations the eigenvectors for the perturbatively corrected eigenvalues are:
\begin{eqnarray}
 \mathbf{\Lambda}^{(k)}_{(0)} &=& \left(
\begin{matrix}
 -\frac{b_1ik}{\tilde\sigma_cT_0}\\
 -\frac{ik}{\tilde\sigma_c}\\
 1
\end{matrix}
\right) , ~~~ \lambda^{(k)}_{(0)} = \frac{k^2}{\tilde\sigma_c}\label{d7}\\
 \mathbf{\Lambda}^{(k)}_{(+)} &=& \left(
\begin{matrix}
 1- \frac{\lambda^{(0)}_{(-)}}{\lambda^{(0)}_{(+)}}\frac{k^2}{\Delta\lambda^{(0)2}}\\
 \frac{\tilde\sigma_c T_0}{b_1\lambda^{(0)}_{(+)}}\left[ 1 - \frac{k^2}{\Delta\lambda^{(0)2}}\right]\\
 \frac{\tilde\sigma_c T_0}{b_1\lambda^{(0)2}_{(+)}}ik
\end{matrix}
\right) , ~~~ \lambda^{(k)}_{(+)} = \lambda^{(0)}_{(+)}+\frac{\lambda^{(0)}_{(-)}k^2}{\lambda^{(0)2}_{(+)}-\lambda^{(0)}_{(+)}\lambda^{(0)}_{(-)}}
\label{d8}\\
\mathbf{\Lambda}^{(k)}_{(-)} &=& \left(
\begin{matrix}
 1 - \frac{\lambda^{(0)}_{(+)}}{\lambda^{(0)}_{(-)}}\frac{k^2}{\Delta\lambda^{(0)2}}\\
 \frac{\tilde\sigma_c T_0}{b_1\lambda^{(0)}_{(-)}}\left[ 1 - \frac{k^2}{\Delta\lambda^{(0)2}}\right]\\
 \frac{\tilde\sigma_c T_0}{b_1\lambda^{(0)2}_{(-)}}ik
\end{matrix}
\right) , ~~~\lambda^{(k)}_{(-)} = \lambda^{(0)}_{(-)}+\frac{\lambda^{(0)}_{(+)}k^2}{\lambda^{(0)2}_{(-)}-\lambda^{(0)}_{(+)}\lambda^{(0)}_{(-)}}
\label{d9}
\end{eqnarray}

\section{Solving for $F^3$ }\label{apc}
Explicitly we have
\begin{eqnarray}
F^{3} &=& \left[\left(\left[2\Xi\right]^{-1}\right)^3_1 - \frac{\Lambda^3_0\Pi^0_1}{2\lambda_0}\right]K^1 
+\frac{\Lambda^3_0\Pi^0_1}{s+2\lambda_0} K^1 \nonumber\\
&+& \left[\left(\left[2\Xi\right]^{-1}\right)^3_3 - \frac{\Lambda^3_0\Pi^0_3}{2\lambda_0}\right]K^3
+\frac{\Lambda^3_0\Pi^0_3}{s+2\lambda_0} K^3 \nonumber\\
&+& \left[\left(\left[2\Xi\right]^{-1}\right)^3_2 - \frac{\Lambda^3_0\Pi^0_2}{2\lambda_0}\right] K^2 
+ \frac{\Lambda^3_0\Pi^0_2}{s+2\lambda_0} K^2
\label{v36}
\end{eqnarray}
Replacing the different expressions we obtain
\begin{eqnarray}
F^{3} &\approx& \left[\left(\left[2\Xi\right]^{-1}\right)^3_1 - \frac{\Lambda^3_0\Pi^0_1}{2\lambda_0}\right]\left\{
\frac{-2}{2A_0+1}\left[\left( \left( \left[2\Xi\right]^{-1}\right)^1_2 -\frac{\Lambda^1_{(0)}\Pi^{(0)}_2}{2\lambda_{(0)}}\right)K^{2}
\right.\right.\nonumber\\
&+&\left.\left.\left( \left( \left[2\Xi\right]^{-1}\right)^1_3 -\frac{\Lambda^1_{(0)}\Pi^{(0)}_3}{2\lambda_{(0)}}\right)K^3\right]
\left(s+2\lambda_{(0)}\right)\right.\nonumber\\
&-&\left.\frac{1}{A_{(0)}}\left[\Lambda^1_{(0)}\Pi^{(0)}_2K^{2}+\Lambda^1_{(0)}\Pi^{(0)}_3K^{3}\right]
+\frac{K^{0}}{\left(s+2\lambda_{0}\right)^{2A_{(0)}}}\right\}\nonumber\\
&+&\frac{\Lambda^3_0\Pi^0_1}{\left(s+2\lambda_0\right)}\left\{
\frac{-2}{2A_0+1}\left[\left( \left( \left[2\Xi\right]^{-1}\right)^1_2 -\frac{\Lambda^1_{(0)}\Pi^{(0)}_2}{2\lambda_{(0)}}\right)K^{2}
\right.\right.\nonumber\\
&+&\left.\left.\left( \left( \left[2\Xi\right]^{-1}\right)^1_3 -\frac{\Lambda^1_{(0)}\Pi^{(0)}_3}{2\lambda_{(0)}}\right)K^3\right]
\left(s+2\lambda_{(0)}\right)\right.\nonumber\\
&-&\left.\frac{1}{A_{(0)}}\left[\Lambda^1_{(0)}\Pi^{(0)}_2K^{2}+\Lambda^1_{(0)}\Pi^{(0)}_3K^{3}\right]
+\frac{K^{0}}{\left(s+2\lambda_{0}\right)^{2A_{(0)}}}\right\}
\nonumber\\
&+& \left[\left(\left[2\Xi\right]^{-1}\right)^3_3 - \frac{\Lambda^3_0\Pi^0_3}{2\lambda_0}\right]K^3
+\frac{\Lambda^3_0\Pi^0_3}{\left(s+2\lambda_0\right)} K^3
+\left[\left(\left[2\Xi\right]^{-1}\right)^3_2- \frac{\Lambda^3_0\Pi^0_2}{2\lambda_0}\right] K^2 \nonumber\\
&+& \frac{\Lambda^3_0\Pi^0_2}{\left(s+2\lambda_0\right)} K^2
\label{v36-b}
\end{eqnarray}

To find the corresponding time dependent function, observe that we can write
\begin{equation}
\int_0^{\infty} du~u^{\alpha}~e^{-\left( s+2\lambda_{(0)}\right)u} = 
\frac{\Gamma\left(\alpha+1\right)}{\left( s+2\lambda_{(0)}\right)^{\alpha+1}}\label{v37}
\end{equation} 
If $\alpha \rightarrow -n$ the integral diverges for $u\rightarrow 0$. But as we are interested in the late behavior of the
fields it is legitimate to compute the limit when $\alpha \rightarrow -n$ and discard the divergent term (that corresponds to
times out of the interval of validity of the approximations made in this paragraph). This we do by adding an 'infrared'
cut-off. We then have
\begin{equation}
\mathcal{J}_n(s) \equiv \int_0^{\infty} du~u^{-n+\epsilon}~e^{-\left( s+2\lambda_{(0)}\right)u} = 
\Gamma\left(1-n+\epsilon\right)\left( s+2\lambda_{(0)}\right)^{n-1-\epsilon}\label{v38}
\end{equation} 
where $\epsilon$ a small parameter. Developing in Laurent series around the pole we have
\begin{eqnarray}
\mathcal{J}_n(s) &\simeq & \frac{\left( -1\right)^n}{n!}\left[ 1 + \epsilon 
\ln\left(\frac{s+2\lambda_{(0)}}{\mu}\right)\right]
\left[\frac{1}{\epsilon} + \psi\left( n+1\right)\right] \left(s+2\lambda_{(0)}\right)^{n-1} \nonumber\\
&\simeq& \frac{\left( -1\right)^n}{n!}\left[\ln\left(\frac{s+\lambda_{(0)}}{\mu}\right)+ \psi\left( n+1\right)\right]
\left(s+2\lambda_{(0)}\right)^{n-1}
\label{v39}
\end{eqnarray}
with $\mu$ a renormalization constant and $\psi = \Gamma^{\prime}/\Gamma$. Finally, for $s\rightarrow 0$ we have
\begin{equation}
\int_0^{\infty} du~u^{-n+\epsilon}~e^{-\left( s+2\lambda_{(0)}\right)u} \rightarrow
\frac{\left( -1\right)^n}{n!}\left[\ln\left(\frac{2\lambda_{(0)}}{\mu}\right)+ \psi\left( n+1\right)\right]
\left( s+2\lambda_{(0)}\right)^{n-1}\label{v40}
\end{equation}
To apply this result to eq. (\ref{v36-b}) we observe that $A_{(0)}\simeq b_2\tilde\tau k^2/\tilde\sigma_c \ll 1$,
and therefore can be discarded in front of 1. The Laplace anti-transformed different terms that appear in expr. 
(\ref{v36-b}) can then be approximated as
\begin{eqnarray}
\left( s+2\lambda_{(0)}\right)^{-1} &\rightarrow & \left[\ln\left(\frac{\lambda_{(0)}}{\mu}\right)+ 
\psi\left(1\right)\right]^{-1} e^{-2\lambda_{(0)}u} \label{v41-b}\\
\left( s+2\lambda_{(0)}\right) &\rightarrow & \frac{2}{u^2}\left[\ln\left(\frac{2\lambda_{(0)}}{\mu}\right)+ 
\psi\left(3\right)\right]^{-1}e^{-2\lambda_{(0)}u}\label{v42}\\
const &\rightarrow & - \frac{1}{u}\left[\ln\left(\frac{2\lambda_{(0)}}{\mu}\right)+ \psi\left(2\right)\right]^{-1}
e^{-2\lambda_{(0)}u}\label{v43}
\end{eqnarray}
and we see that the contribution that gives the slower decay comes from correspondence (\ref{v41-b}). 
We then keep only those terms, obtaining
\begin{eqnarray}
F^{3} &\simeq & -\frac{\Lambda^3_0\Pi^0_1}{\left(s+2\lambda_0\right)}
\frac{1}{A_{(0)}}\left[\Lambda^1_{(0)}\Pi^{(0)}_2K^{2}+\Lambda^1_{(0)}\Pi^{(0)}_3K^{3}\right]+
\frac{\Lambda^3_0\Pi^0_1}{\left(s+2\lambda_0\right)^{1+2A_{(0)}}}K^{0}
 \nonumber\\
&+&\frac{\Lambda^3_0\Pi^0_3}{\left(s+2\lambda_0\right)} K^3
 + \frac{\Lambda^3_0\Pi^0_2}{\left(s+2\lambda_0\right)} K^2
\label{v433}
\end{eqnarray}
The calculation of elements $\Pi^i_j$ is rather long but straightforward, here we quote the one in the term
with $K^{0}$ as this term gives the main contribution. It reads $\Pi^0_1 = ikb_2\tilde\tau T_0/b_1$, and 
we then have
\begin{equation}
F^{3} \simeq  
\frac{b_2\tilde\tau T_0 K^{0}}{b_1\left(s+2\lambda_0\right)^{1+2A_{(0)}}} ik
\label{v44}
\end{equation}
Leaving aside the constant factor in expr. (\ref{v41-b}) the corresponding anti-transformed function is
\begin{eqnarray}
B_y^{(k)}(u) &\sim & \left[\frac{b_2\tilde\tau T_0 K^{0}}{b_1} ik u^{2A_{(0)}} + \mathcal{O}\left(\frac{1}{u}\right)\right]
\exp \left[-2\lambda_0 u\right]
\label{v45}
\end{eqnarray}

\end{document}